\documentclass{aastex631}

\usepackage{tabularx}
\usepackage{graphicx}

\newcommand{\um}{$\mu$m}

\newcommand{\rev}{}

\received{2025 January 15}
\revised{2025 April 7}
\accepted{2025 April 8}
\submitjournal{The Astronomical Journal}

\begin{document}

\title{The Milky Way Project MOBStIRS: Parametrizing Infrared Stellar-Wind Bow Shock Morphologies with Citizen Science 
}


\author{Angelica S. Whisnant}
\affiliation{Department of Physics and Astronomy, California State Polytechnic University Pomona, 3801 West Temple Avenue, Pomona, CA 91768, USA}
\affiliation{Department of Astronomy, The Ohio State University, 140 West 18th Avenue, Columbus, OH 43210, USA}

\author{Matthew S. Povich}
\affiliation{Department of Physics and Astronomy, California State Polytechnic University Pomona, 3801 West Temple Avenue, Pomona, CA 91768, USA}

\author{Nikhil Patten}
\affiliation{Department of Physics and Astronomy, University of Wyoming, Dept 3905, Laramie, WY 82070-1000, USA}

\author{Henry A. Kobulnicky}
\affiliation{Department of Physics and Astronomy, University of Wyoming, Dept 3905, Laramie, WY 82070-1000, USA}



\begin{abstract}
 Mass-loss influences stellar evolution, especially for massive stars with strong winds. 
Stellar wind bow shock nebulae driven by Galactic OB stars can be used to measure mass-loss rates ($\dot{M}$). 
The standoff distance ($R_{0}$) between the star and the bow shock is set by momentum flux balance between the stellar wind and the surrounding interstellar medium (ISM). We created the Milky Way Project: MOBStIRS (Mass-loss rates for OB Stars driving IR bow Shocks) using the online Zooniverse citizen science platform. We enlisted several hundred students to measure $R_0$ and two other projected shape parameters for 764 cataloged IR bow shocks. MOBStIRS incorporated 1528 
JPEG cutout images produced from Spitzer GLIMPSE and MIPSGAL survey data. 
 Measurements were aggregated to compute shape parameters for each bow shock image deemed high-quality by participants. The average statistical uncertainty on $R_0$ is 12.5\% but varies from ${<}5\%$ to ${\sim}40\%$ among individual bow shocks, contributing significantly to the total error budget of $\dot{M}$. The derived nebular morphologies agree well with \rev{(magneto)}hydrodynamic simulations of  bow shocks driven by the winds of OB stars moving at $V_a = 10$--40~km~s$^{-1}$ with respect to the ambient interstellar medium (ISM). A systematic correction to $R_0$ to account for viewing angle appears unnecessary for computing $\dot{M}$. Slightly more than half of MOBStIRS bow shocks are asymmetric, \rev{which could indicate} anisotropic stellar winds, ISM clumping on sub-pc scales, \rev{time-dependent instabilities, and/or misalignments between the local ISM magnetic field and the star-bow shock axis.}




\end{abstract}

\keywords{Stellar bow shocks(1586) --- Educational software(1870) --- Infrared astronomy(786) --- Massive stars(732) --- Stellar mass loss(1613) --- Circumstellar dust(236)}


\section{Introduction} \label{sec:intro}
Massive stars lose mass through their stellar winds, the radiation-driven outflow of ionized plasma from the star itself \citep{lucysolomon70, castor1975, pauldrach1986}.
These mass-loss rates significantly impact the evolution and end states (e.g., neutron stars versus black hole remnants) of massive O- and early B-type stars \citep{smith2014}. They also deposit significant amounts of energy into the ISM, which is key for galaxy evolution models \citep{hopkins2014, fier2016}. 

However, the mass-loss rates of OB stars historically have been poorly constrained, with various theoretical calculations and observational values inconsistent with each other, exhibiting discrepancies of an order of magnitude or more \citep{vink2001, martins2005, fullerton2006}.
Mass-loss rates are very sensitive to variations in wind density, or clumping  \citep{ebbets1982, fullerton1996}. \citet{brands2022} found significant clumping in their sample of 53 O- and 3 WNh-type stars, and also that lower clumping correlates with higher mass-loss rates.
\citet{bjorklund2021.II, bjorklund2023.III} shows when clumping is considered, computed mass-loss rates are lower (by a factor of $\sim$3 for O stars and by one to two orders of magnitude for B giants/supergiants) than what is implemented in evolution calculations \citep{vink2000, vink2001} . Similar conclusions have been drawn from observational work \citep[e.g.,][]{hawcroft2021}.

Commonly used observational methods to measure mass loss rates include: H$\alpha$ emission \citep{leitherer1988, lamers1993, puls1996, markova2004, martins2005}, radio emission \citep{lamers1993}, X-ray and mid-IR spectra \citep{cohen2014}, and UV P Cygni profiles \citep{garmany1981}. These methods rely on detailed spectroscopy, which can be difficult to obtain, particularly for distant stars suffering high extinction.

Infrared stellar-wind bow shocks (IR bow shocks; \citealt{vanburen1988}) provide an observational approach to measure mass-loss rates for hundreds of candidate OB stars \citep{k16,mwpdr2}, using a method that is insensitive to clumping and far less affected by extinction \citep{kobulnicky2010,masslossmethod,k19}. 
An IR bow shock forms an arc tracing the pressure balance between the stellar wind that is moving supersonically (at speed $V\textsubscript{a}$) relative to the ambient interstellar medium (ISM):
\begin{equation}
    \frac{1}{2} \rho\textsubscript{w} V \textsubscript{w}\textsuperscript{2} = \frac{1}{2} \rho\textsubscript{a} V\textsubscript{a}\textsuperscript{2}
    \label{eqn:pressurebal}
\end{equation}
\citep{wilkin1996,kobulnicky2010},
where $\rho\textsubscript{w}$ is the density of the stellar wind, $V\textsubscript{w}$ is the velocity of the stellar wind, and $\rho\textsubscript{a}$ is the density of the ambient ISM.


Along with the stellar mass-loss rate $\dot{M}$, the ``standoff distance'' $R_0$ between the driving star and the apex of the arc sets the stellar-wind density,
\begin{equation}
    \rho\textsubscript{w} = \frac{\dot{M}}{4\pi R\textsubscript{0} \textsuperscript{2} V\textsubscript{w}}.
    \label{eqn:winddensity}
\end{equation}
Substituting Equation \ref{eqn:winddensity} into \ref{eqn:pressurebal} and solving for mass-loss rate yields
\begin{equation}
    \dot{M} = 4 \pi R\textsubscript{0} \textsuperscript{2} \frac{V\textsubscript{a} \textsuperscript{2} \rho \textsubscript{a}} {V\textsubscript{w}}
    \label{eqn:massloss}
\end{equation}
\citep{masslossmethod,k19}, which depends $R_{0}^2$. The accuracy of measured standoff distances, including statistical and systematic uncertainties, strongly influences the derived  mass-loss rates.

\begin{figure}[ht!]
  \centering
  \begin{minipage}[b]{0.5\textwidth}
    \includegraphics[width=\textwidth]{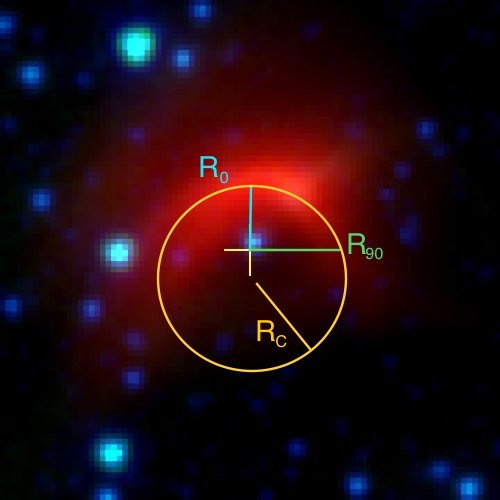}
  \end{minipage}
\caption{Image of an IR bow shock created using 24~\um ~data from MIPSGAL (red) and 8.0~\um~and 4.5~\um~data from GLIMPSE (green and blue, respectively). Overlaid are the IR bow shock shape parameters of interest: $R_0$ (measured from the driving star to the apex of the arc), $R_{90}$ (the distance from the driving star to the arc, perpendicular to the symmetry axis of the bow shock), and $R_c$ (the best fitting circle along the arc of the bow shock).
\label{fig:parameters}}
\end{figure}

In this paper, we provide new and improved measurements of $R_0$ for hundreds of IR bow shocks, including uncertainties.  We also measure several other bow shock size and shape parameters that can be used to constrain the intrinsic 3D nebular morphology and compare to predictions from analytical theory and simulations.
\citet[][hereafter TH18]{truevsapparent} 
developed a general theory for the family of possible shapes for cylindrically-symmetric bow shocks, including the effects of viewing angles on observed arc shapes. \rev{Potential shapes are ``quadrics of revolution" \citep{goldman1983, gfrerrer2009}, wilkinoids (a 3D generalization of the 2-D analytical model developed by \citealt{wilkin1996}), cantoids \citep{canto1996}, and ancantoids (a generalization of the cantoid for varying relative momenta of interacting winds and ISM flows)}.
TH18 parameterized these shapes in terms of $R_0$, $R_{90}$ (the distance from the driving star to the arc wing, perpendicular to the symmetry axis of the bow shock) and $R_c$ (the best fitting circle along the arc of the bow shock, which is not necessarily centered on the driving star). These are illustrated in Figure \ref{fig:parameters}.

We use data from the \textit{Spitzer Space Telescope} to create images of individual IR bow shocks to be measured by citizen scientists through the online Zooniverse platform. 
The Galactic Legacy Infrared Mid-Plane Survey Extraordinaire \citep[GLIMPSE 4.5 and 8.0~\um;][]{glimpse2003, glimpse2009} and MIPSGAL \citep{mipsgal} surveyed the inner Galactic mid-plane with continuous coverage of Galactic longitudes $0\arcdeg\le |l| \le 65\arcdeg$ and latitudes $|b|\le 1.2\arcdeg$. 

Zooniverse has hosted hundreds of successful citizen science projects over the past 15 years, and using citizen scientists for scientific research has been proven to be a reliable method of data collection. The Milky Way Project (MWP) was first introduced in 2010 as the 10th Zooniverse project, and subsequently the $\sim$3 million classifications made by citizen scientists were used identify 2600 candidate IR bubbles and 599 IR bow shock driving stars \citep{mwpdr1, mwpdr2}. About 300 of the IR bow shocks were new discoveries, not previously cataloged by \citet[][hereafter K16]{k16}.
The citizen science approach is well-suited to the large number of images that present a wide variation in morphology, which can be challenging for automated measurement techniques \citep[see][who performed such automated measurements for the K16 bow shocks]{henney4}.  IR bow shocks are observed against complex, spatially-varying nebular background emission, and at varying spatial resolution, given the range of distances and physical sizes. Compared to ``by-hand'' measurements by individual scientists (K16), obtaining repeated measurements per image by multiple citizen scientists mitigates human bias and provides a way to quantify uncertainties.

In Section \ref{sec:method}, we describe the process of preparing the data and creating and administering the citizen science website, as well as data analysis details. In Section \ref{sec:results}, we present our final projected shape parameter values, comparing $R_0$ with a value previously determined by K16 and both sides of $R_{90}$. In Section \ref{sec:discussion} we discuss the implications of our results for theoretical models of bow shocks and $\dot{M}$ determinations, and summarize our conclusions in Section \ref{sec:conclusion}.

\section{Method} \label{sec:method}
\subsection{Image Preparation} \label{subsec:preparation}

We created image cutouts for the MOBStIRS website, each centered on the position of one of 764 IR bow shock driving stars from the K16  or Milky Way Project \citep{mwpdr2} catalogs (the subset located within the GLIMPSE+MIPSGAL survey areas).

Modifying the approach used by \citet{mwpdr2}, we started with wide-field FITS image mosaics produced by the GLIMPSE team, including 24~\um\ mosaics from the MIPSGAL survey. 
We produced JPEG image cutouts zoomed-in, scaled, and rotated to display each IR bow shock in a visually uniform manner (see Figures~\ref{fig:parameters}--\ref{fig:examples} for examples), to make the measurement task of the citizen scientists as consistent as possible. Color channels in all images were assigned as red = 24~\um, green = 8.0~\um, and blue = 4.5~\um, with a square-root stretch function applied independently to each channel in each cutout. Visual examination of the output images and experimentation with the blackpoints revealed that setting the faintest 40\% of pixels to black and the brightest 2\% to white provided optimal view of the structure of bow shock arcs. This represents a stronger suppression of  faint pixels compared to previous MWP versions, which was helpful to suppress diffuse background nebular emission at 8~\um\ and 24~\um\ that tends to obscure the bow shocks.

The cataloged position angles \rev{\citep{k16}} were used to orient each image such that the apex of the arc was at the top, above the driving star. Cataloged $R_0$ values were used to crop and resample the FITS images to $500\times 500$ JPEG pixel dimensions. The zoom was chosen such that the diagonal length of each square image equaled $10R_0$ for $R_0\ge 10\arcsec$, or $20R_0$ otherwise.

To simplify the citizen science task of measuring $R_{90}$, which for real bow shocks can assume different values toward the right or left wings of the arc, two versions of each IR bow shock were made: an “original” version and a “flipped” version mirrored around the axis joining the star to the arc apex. MOBStIRS users were instructed to measure $R_{90}$ toward the right wing only for a given image, whether original or flipped.
Each driving star was marked by a yellow ``+” symbol, to ensure that citizen scientists would use the correct starting point for measuring $R_{0}$ and $R_{90}$. Many images feature several visible stars that could be confused with the driving star, and some original candidate driving star identifications in the K16 or MWP catalogs were subsequently changed based on subsequent re-examination or spectral classification \citep{k19,chick20}.

\subsection{Creating the Website} \label{subsec:website}
\begin{figure}[ht!]
  \centering
  \begin{minipage}[b]{0.8\textwidth}
    \includegraphics[width=\textwidth]{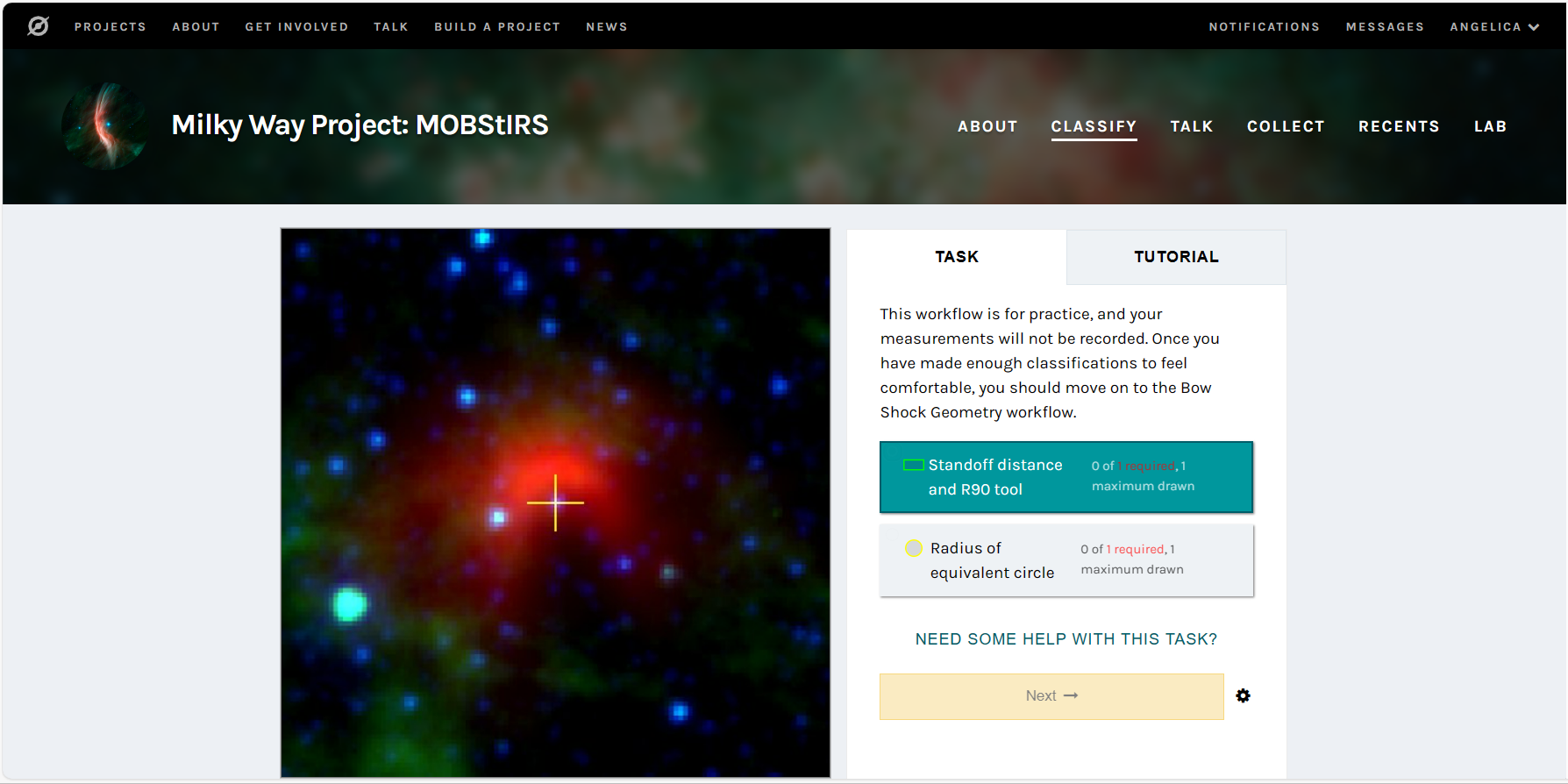}
  \end{minipage}
\caption{Screenshot of the interface used by the citizen scientists to make classification drawings on each image.
\label{fig:interface}}
\end{figure}

We used the \href{https://www.zooniverse.org/lab}{Zooniverse project builder} to create the MOBStIRS website hosting our citizen science project.
The MOBStIRS site was not made public because the number of bow shock images was relatively small compared to the tens of thousands of data subjects that public Zooniverse projects, including MWP DR1 and DR2, typically require. 
The MOBStIRS site otherwise incorporated many elements found among public Zooniverse projects, including background information on the project (``About'' tab),  features allowing participants to create collections of favorite images and discuss these and other aspects of the project with each other on discussion boards, and a tutorial  explaining how to perform each measurement, with image and video examples. Tabs and links to these features are visible in Figure~\ref{fig:interface}, which shows the MOBStIRS classification interface.

We uploaded 1528 JPEG images (from the original \textit{Spitzer} GLIMPSE and {MIPSGAL} survey mosaics) to the MOBStIRS ``bow shock geometry'' workflow, which we had set up using the Zooniverse Project Builder. A subset of 180 bow shock driving stars were identified as high priority for ground-based spectroscopic followup, so we separately uploaded 360 duplicate JPEG images (both original and flipped versions) for these objects, giving us the ability to reduce the MOBStIRS set to only this set for certain time periods, which produced higher classification counts for these objects. 

Zooniverse requires a ``manifest,'' a CSV file containing a table of metadata for all uploaded image subjects in a given batch. The MOBStIRS manifests included central Galactic coordinates and plate scales so that citizen science measurements in image pixels could be converted back to sky coordinates.
The Zooniverse platform assigns a ``subject ID" to each image and displays these subjects at random for the citizen scientists to measure at the user interface. 

\begin{figure}[ht!]
  \centering
  \begin{minipage}[b]{0.47\textwidth}
    \includegraphics[width=\textwidth]{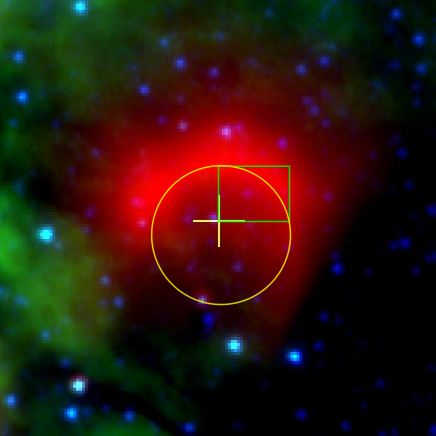}
  \end{minipage}
  \hfill
  \begin{minipage}[b]{0.47\textwidth}
    \includegraphics[width=\textwidth]{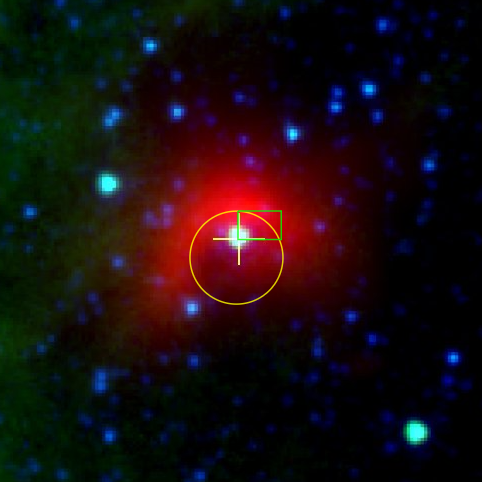}
  \end{minipage}

\caption{Example MOBStIRS JPEG image cutouts of IR bow shocks created using 24~\um\ data from MIPSGAL (red) and 8.0~\um~and 4.5~\um~from GLIMPSE (green and blue, respectively). Overlays show reasonable classification drawings using the Zooniverse Box (green) and Circle (yellow) tools. These examples were included in the tutorial.
\label{fig:examples}}
\end{figure}

We chose two drawing tools from among the standard options in the project builder by which users could make the three required measurements, the Box tool (to measure values for R\textsubscript{0} and R\textsubscript{90} as the height and width, respectively, of a rectangle) and the Circle tool to measure R\textsubscript{c} (Figure \ref{fig:examples}). The Box tool does not allow for rotation, which helped motivate our decision to rotate the images to the common orientation (e.g., Figures~\ref{fig:interface} and \ref{fig:examples}). We added a final question task asking: ``Was this image of sufficient quality that you could make reasonably precise and accurate drawings?" The citizen scientists responded to this by clicking ``Yes" or ``Not really." In the tutorial slides, citizen scientists were instructed to always select ``Not really" when the apex was not directly above the star or when the wing to the right of the star was not visible.
 Participants were instructed to begin with the ``training set" workflow, which contained a set of 24 relatively well-defined IR bow shock images, as practice before continuing to the bow shock geometry workflow. Classifications made using the training workflow were excluded from our analysis.

MOBStIRS was presented to undergraduate students enrolled in astronomy courses. The presentation included the scientific background of the project, instructions on navigating the website, and instructions on how to make the measurements. As students began practicing with the training set, we encouraged them to ask questions about the measurements. Once they were comfortable making measurements on their own, they were tasked to make measurements in the main workflow.  Students in the Cal Poly Pomona courses were incentivized to complete measurements by giving course credit upon completion of at least 50 images, including tagging at least one ``favorite'' image for discussion among their classmates. The first few classes engaged with the site during the development process completed a separate online survey that requested feedback on the usability and documentation. This feedback was used to improve aspects of the website, namely the ``About" page and the tutorial, to ensure students were able to understand the astronomy concepts involved in the project and the measurements expected from the content of the website. Two large-enrollment General Education classes completed a worksheet that we developed to accompany MOBStIRS and enhance its pedagogical value. The worksheet contained qualitative and quantitative questions designed to assess students' understanding of massive stars and mass-loss rates in the the context of stellar evolution and Galactic stellar populations.

In Summer 2023 eight undergraduates participating in the University of Wyoming Astronomy Research Experiences for Undergraduates program  completed a large set of classifications on the priority bow shock image sets. These students were given the same training as the classroom cohorts, and driven by their pre-existing enthusiasm for astronomy and greater background knowledge than undergraduate non-majors, they were willing and able to make more classifications more quickly.

\subsection{Data Analysis} \label{subsec:analysis}

Classification data were 
exported from the MOBStIRS website as CSV files. These contained the name and coordinates of each classified IR bow shock, the location and size (in pixels) of the shapes drawn on each image, a flag indicating whether the image was the original or flipped version, and the plate scale converting image pixels to arcseconds.
Zooniverse provides aggregation code called \texttt{panoptes} \citep{zooniverse_agg} to process the raw classification data. We used the question extractor and reducer, rectangle extractor and reducer, and shape extractor and reducer, which produced three CSV files of measurement data for each subject, one for each of the tools used (rectangle, circle, question). These CSV files served as inputs to a custom  analysis pipeline we developed using Python. Because some subjects were uploaded multiple times and were therefore associated with more than one subject ID, we combined the sets of data that were associated with one subject, referencing the common subject name, which was based on the central Galactic coordinates of each image. 
We determined the total number of measurements we had for each subject by adding the number of ``yes" and ``not really" votes received on confidence, then removed subjects that had fewer than 5 measurements. This left us with 1494 subjects. We then removed subjects that had more ``not really" than ``yes" votes for confidence level, leaving us with 995 subjects. For each of the parameters measured, we rejected ${>}1$-$\sigma$ outliers and then computed the mean and standard deviation of the remaining data using one iteration of \texttt{astropy.stats.sigma\_clipped\_stats}. We then cross-referenced the original and flipped versions of each bow shock (also by referencing the subject name), leaving us with 586 unique IR bow shocks with measurements. We converted the classification data from pixels to arcseconds using the conversion factors from the manifest. We determined final values for the projected shape parameters $R_0$, $R_{90}$, and $R_c$ for each IR bow shock that had data for both original and flipped image versions by taking the weighted average based on the number of measurements made. Uncertainties for these bow shocks were added in quadrature. 


\section{Results} \label{sec:results}

Participants in MOBStIRS included more than 250 students from 5 astronomy courses at Cal Poly Pomona, a class at the University of North Texas, and 8 students interns in the University of Wyoming Research Experiences for Undergraduates summer 2023 cohort, making 17,300 total classifications. Table \ref{tab:cs_num_measurements} 
lists the participants by number of classifications completed.

Table \ref{tab:data} presents the results of all measurements made by MOBStIRS citizen scientists for the 586 IR bow shocks with a majority of users reporting they were confident in their measurements (Frac. Yes $>0.5$).  
The total number of citizen science classifications for each IR bow shock, $N_{CS}$, is irrespective of a user's response to the question about subjective measurement confidence. A large majority of MOBStIRS bow shocks have $N_{CS}>10$. 
Two values for standoff distance are included: $R_0$(pre) and $R_0$(CS) are the previous measurement and citizen science measurements, respectively. The next 5 columns present uncertainty on standoff distance ($\sigma_{R_0}$), and the values of the other two shape parameters and associated uncertainties measured from MOBStIRS. Averaged across the entire sample, the percentage uncertainty for each of the projected shape parameters is $\langle\sigma_{R_0}/R_0\rangle=12.5\%$, $\langle \sigma_{R_{90}}/R_{90}\rangle= 14.5\%$, and $\langle \sigma_{R_c}/R_c\rangle = 12.4\%$.

The O+F column provides a flag indicating whether the IR bow shock had successful classifications on both the original and flipped image versions (O+F = 2) or only one of these (O+F = 1). A related flag, Asym., is set to 1 if the bow shock was determined to be asymmetric, 0 if symmetric, or -99 if O+F = 1 (see Section~\ref{sec:asymm} below).

The last 4 columns contain the projected planitude ($\Pi' \equiv R_c/R_0$) and projected alatude ($\Lambda' \equiv R_{90}/R_0$) and their uncertainties, calculated from the MOBStIRS data (these parameters are defined in TH18).







\begin{deluxetable*}{cccc}
\tablecaption{List of citizen scientist participants by unsername and number of images classified}
\tablewidth{0pt}
\tabletypesize{\tiny}
\label{tab:cs_num_measurements}

\tablehead{
\colhead{Username} 
& \colhead{\# classifications} 
& \colhead{Username} 
& \colhead{\# classifications} 
}

\startdata
arosenthal33 & 364 & JJUEOM08 & 189 \\
\\
Dthope & 361 & Griffin3m & 162 \\
\\
Will5221 & 361 & NoahScaletta & 154 \\
\\
sandrews82 & 360 & beduron & 146 \\
\\
evelynnp & 234 & ThallisUW & 146\\
\\
\enddata
\tablecomments{A portion of this table, presenting the 10 citizen scientists with the largest classification counts, is provided here to illustrate its format and content. The full table is available in its entirely in machine-readable form.}

\end{deluxetable*}
\begin{deluxetable*}{ccccccccccccccccc}
\tablecaption{Apparent shape parameter measurements for IR bow shocks in the MOBStIRS final sample, including uncertainties. 
 }
\tablewidth{0pt}
\tabletypesize{\tiny}
\label{tab:data}

\tablehead{
\colhead{Bowshock \# $^a$} 
& \colhead{Subject Name} 
& \colhead{$N_{\rm CS}$} 
& \colhead{Frac. Yes} 
& \colhead{$R_0$(pre)} 
& \colhead{$R_0$(CS)} 
& \colhead{$\sigma_{R_0}$} 
& \colhead{$R_{90}$} 
& \colhead{$\sigma_{R_{90}}$} 
& \colhead{$R_c$} 
& \colhead{$\sigma_{R_c}$} 
& \colhead{O+F} 
& \colhead{Asym.} 
& \colhead{$\Pi'$} 
& \colhead{$\sigma_{\Pi'}$} 
& \colhead{$\Lambda'$} 
& \colhead{$\sigma_{\Lambda'}$} 
\\
\colhead{} 
& \colhead{} 
& \colhead{} 
& \colhead{} 
& \colhead{arcsec} 
& \colhead{arcsec} 
& \colhead{arcsec} 
& \colhead{arcsec} 
& \colhead{arcsec} 
& \colhead{arcsec} 
& \colhead{arcsec} 
& \colhead{} 
& \colhead{} 
& \colhead{} 
& \colhead{} 
& \colhead{} 
& \colhead{} 
}

\startdata
26 & G011.0709-00.5437 & 56 & 0.93 & 8.2 & 8.6 & 0.76 & 11.55 & 1.01 & 12.27 & 1.39 & 2 & 0 & 1.43 & 0.21 & 1.34 & 0.17 \\
\\
50 & G014.6719-00.4770 & 48 & 0.77 & 7.2 & 9.32 & 1.55 & 14.34 & 1.76 & 16.4 & 2.15 & 2 & 1 & 1.76 & 0.37 & 1.54 & 0.32 \\
\\
100 & G023.0958+00.4411 & 48 & 0.79 & 6.7 & 7.18 & 0.65 & 8.36 & 1.54 & 8.69 & 1.35 & 2 & 0 & 1.21 & 0.22 & 1.16 & 0.24 \\
\\
69 & G017.2748-00.4885 & 46 & 0.63 & 7.6 & 7.76 & 1.09 & 10.93 & 1.75 & 13.31 & 2.09 & 2 & 0 & 1.71 & 0.36 & 1.41 & 0.3 \\
\\
282 & G053.4178+00.0990 & 44 & 0.66 & 11.3 & 14.32 & 1.34 & 18.05 & 3.22 & 20.7 & 2.04 & 2 & 1 & 1.45 & 0.2 & 1.26 & 0.25 \\
\\
748 & MWP2G030.6628-00.08784 & 43 & 0.91 & 8.4 & 9.07 & 1.04 & 15.82 & 1.16 & 18.94 & 1.66 & 2 & 0 & 2.09 & 0.3 & 1.74 & 0.24 \\
\\
28 & G011.6548+00.4943 & 42 & 0.86 & 13.4 & 13.79 & 0.9 & 23.56 & 1.45 & 29.1 & 3.06 & 2 & 0 & 2.11 & 0.26 & 1.71 & 0.15 \\
\\
165 & G032.0177-00.4999 & 42 & 0.64 & 8.6 & 9.14 & 1.18 & 12.97 & 1.56 & 13.87 & 1.64 & 2 & 0 & 1.52 & 0.27 & 1.42 & 0.25 \\
\\
464 & G308.0703+00.2120 & 42 & 0.93 & 12.4 & 12.32 & 1.24 & 18.42 & 1.47 & 19.44 & 1.67 & 2 & 0 & 1.58 & 0.21 & 1.5 & 0.19 \\
\\
18 & G008.3690+00.0239 & 41 & 0.98 & 27 & 25.49 & 1.17 & 40.2 & 2.72 & 44.27 & 3.08 & 2 & 0 & 1.74 & 0.14 & 1.58 & 0.13 \\
\\
\enddata
\tablecomments{A portion of this table, presenting the 10 bow shocks with the largest MOBStIRS classification count, is provided here to illustrate its format and content. The full table is available in its entirely in machine-readable form.\\
$^a$ These numbers were assigned according to \citet{k16} and \citet{mwpdr2}.}
\end{deluxetable*}
\subsection{Standoff Distance Comparisons}\label{sec:r0comp}

\begin{figure}[ht!]
  \centering
  \begin{minipage}[b]{0.49\textwidth}
    \includegraphics[width=\textwidth]{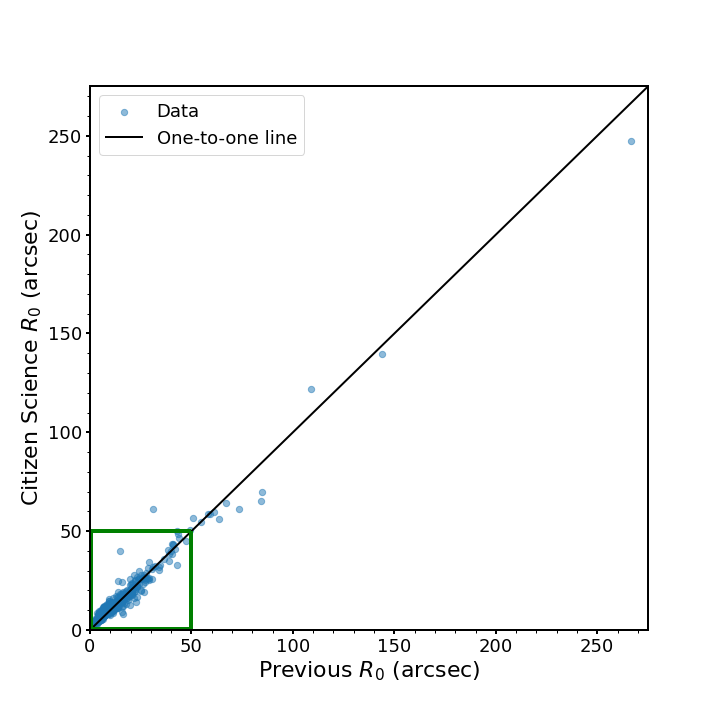}
  \end{minipage}
  \hfill
  \begin{minipage}[b]{0.49\textwidth}
    \includegraphics[width=\textwidth]{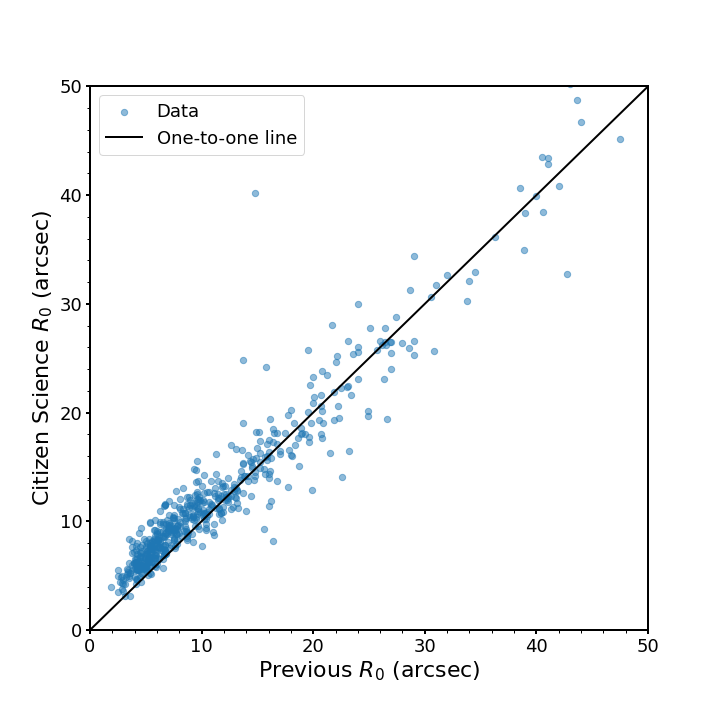}
  \end{minipage}
    
\caption{Plots comparing the $R_0$ value determined by a previous ``expert" scientist ($x$-axes) with the value determined by the citizen scientists ($y$-axes). The same data are plotted twice, first with axes spanning the full range of data, with a green box showing the area zoomed in for the second plot. 
\label{fig:r0comp}}
\end{figure}

In Figure \ref{fig:r0comp} we compare the values for $R_0$ from MOBStIRS to the values previously measured by a few ``expert" scientists using the original GLIMPSE and MIPSGAL FITS mosaics \citep{k16}. 
We find that for smaller IR bow shocks (Previous $R_0\la 10\arcsec$), citizen science measurements of $R_0$ are systematically larger, by about 1.5\arcsec, than the previous measurements. 
This systematic offset is comparable to the pixel size (1.2\arcsec) of the original FITS mosaics, and smaller than the 5.5\arcsec\ diffraction-limited resolution of the MIPS 24~\um\ bandpass. The affected bow shocks are hence not well-resolved and more challenging to measure accurately. The offset occurs below the size cutoff for which we changed the zoom factor for the MOBStIRS images, suggesting it could be due, in part, to smaller apparent sizes of the bow shock nebulae within the images presented for measurements. \citet{henney4} measured $R_0$ for each of 471 K16 bow shocks within the MIPSGAL survey (a large subset of the MOBStIRS sample) using an automated fitting process, and their fitted $R_0$ was also systematically larger than the K16 cataloged values for small bow shocks.


\begin{figure}[ht!]
  \centering
    \includegraphics[width=0.42\textwidth]{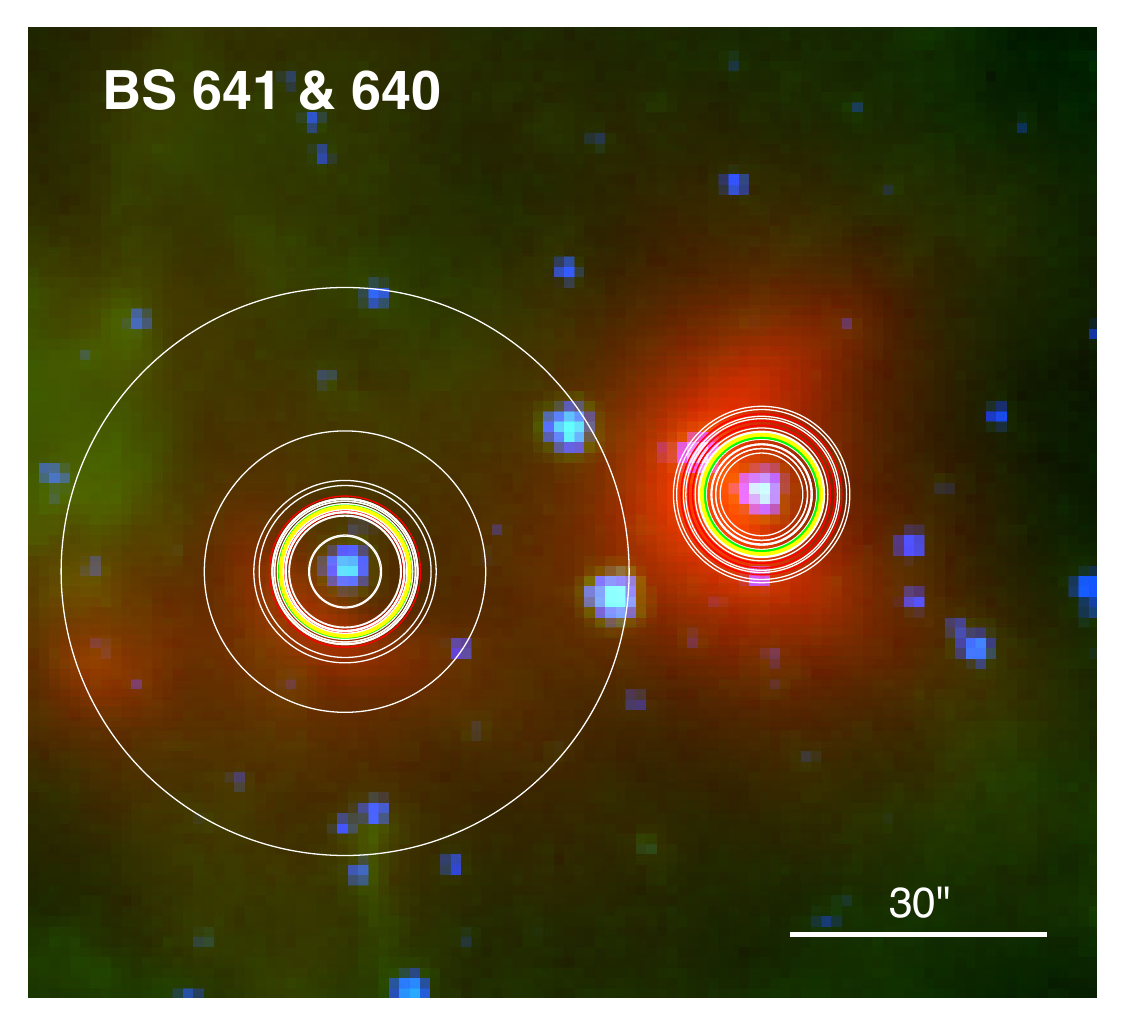}
    \includegraphics[width=0.42\textwidth]{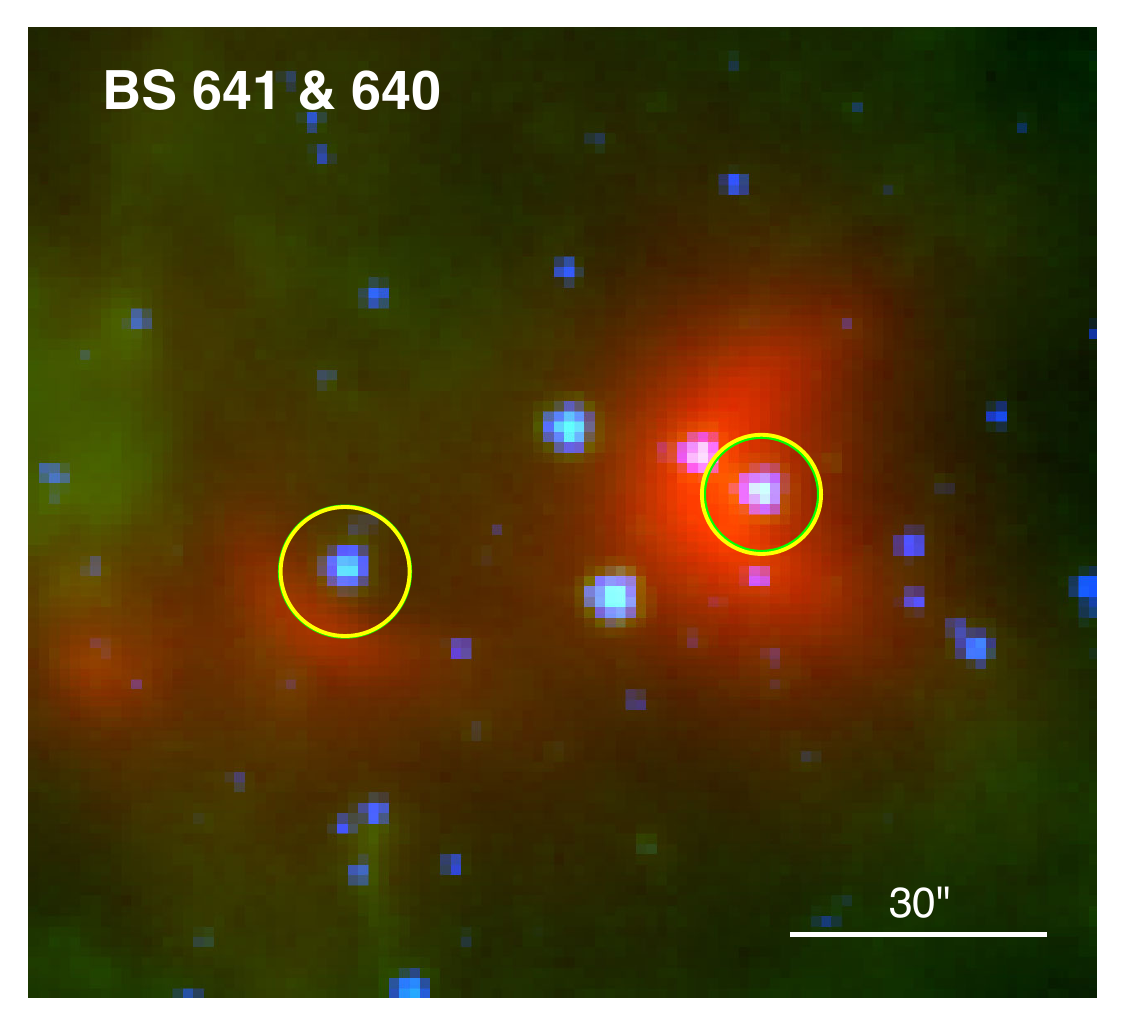} \\
    \includegraphics[width=0.42\textwidth]{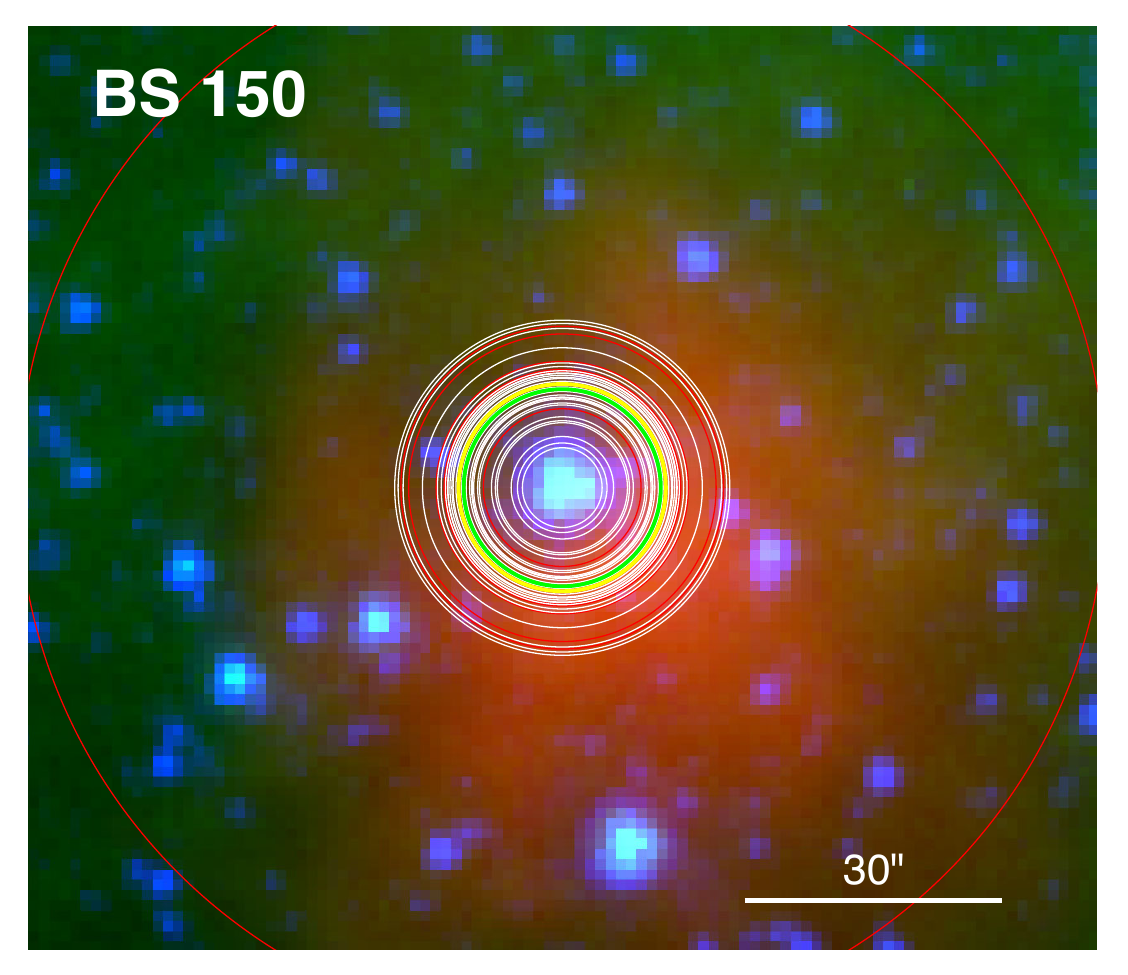}
    \includegraphics[width=0.42\textwidth]{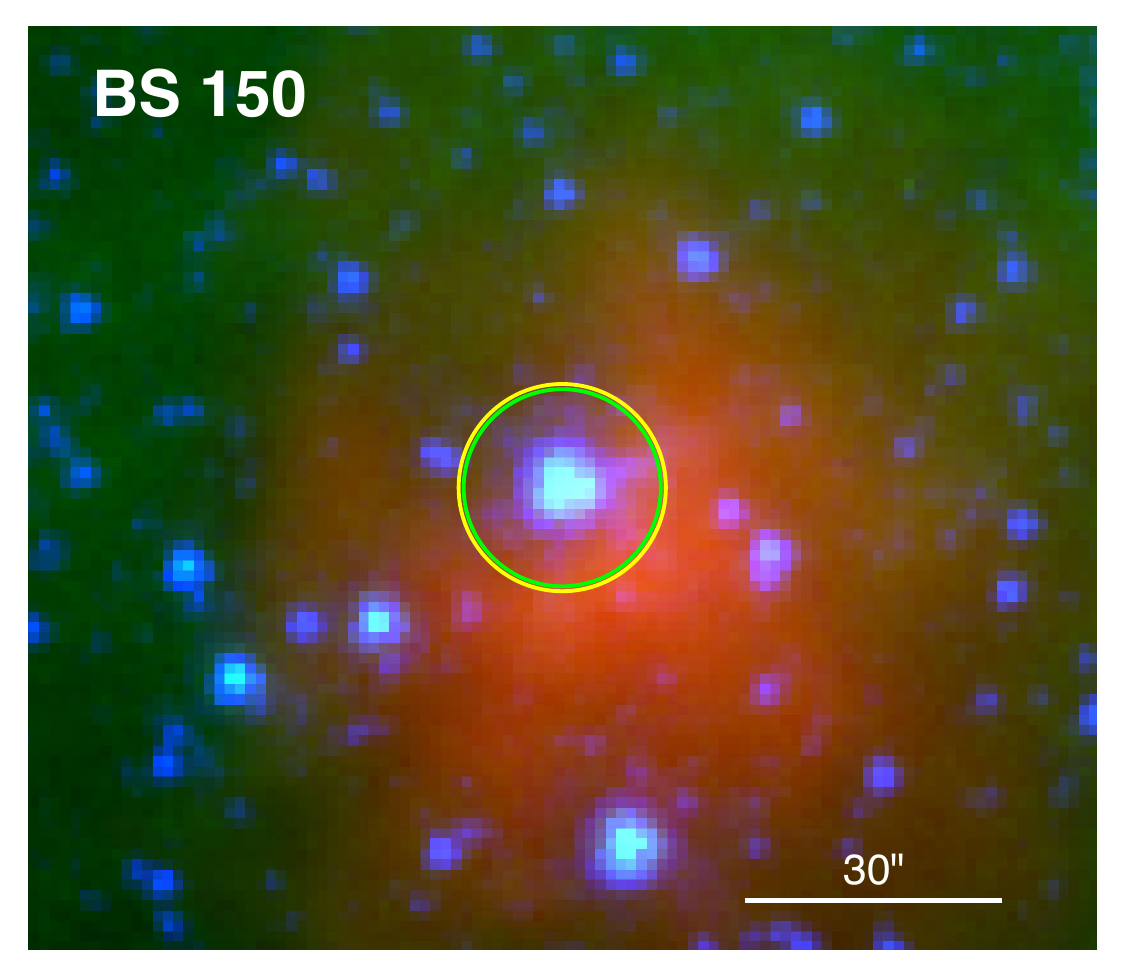} \\
    \includegraphics[width=0.42\textwidth]{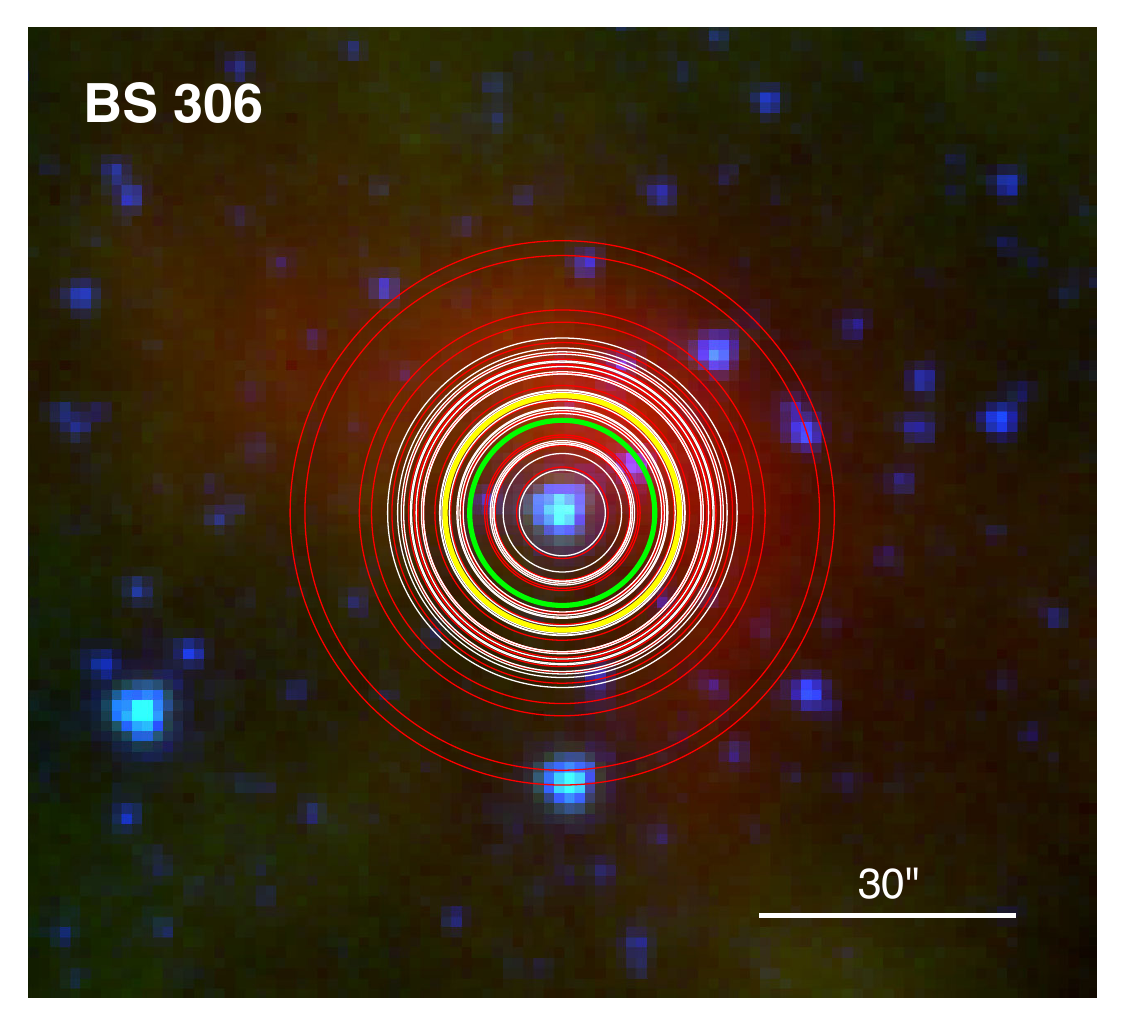}
    \includegraphics[width=0.42\textwidth]{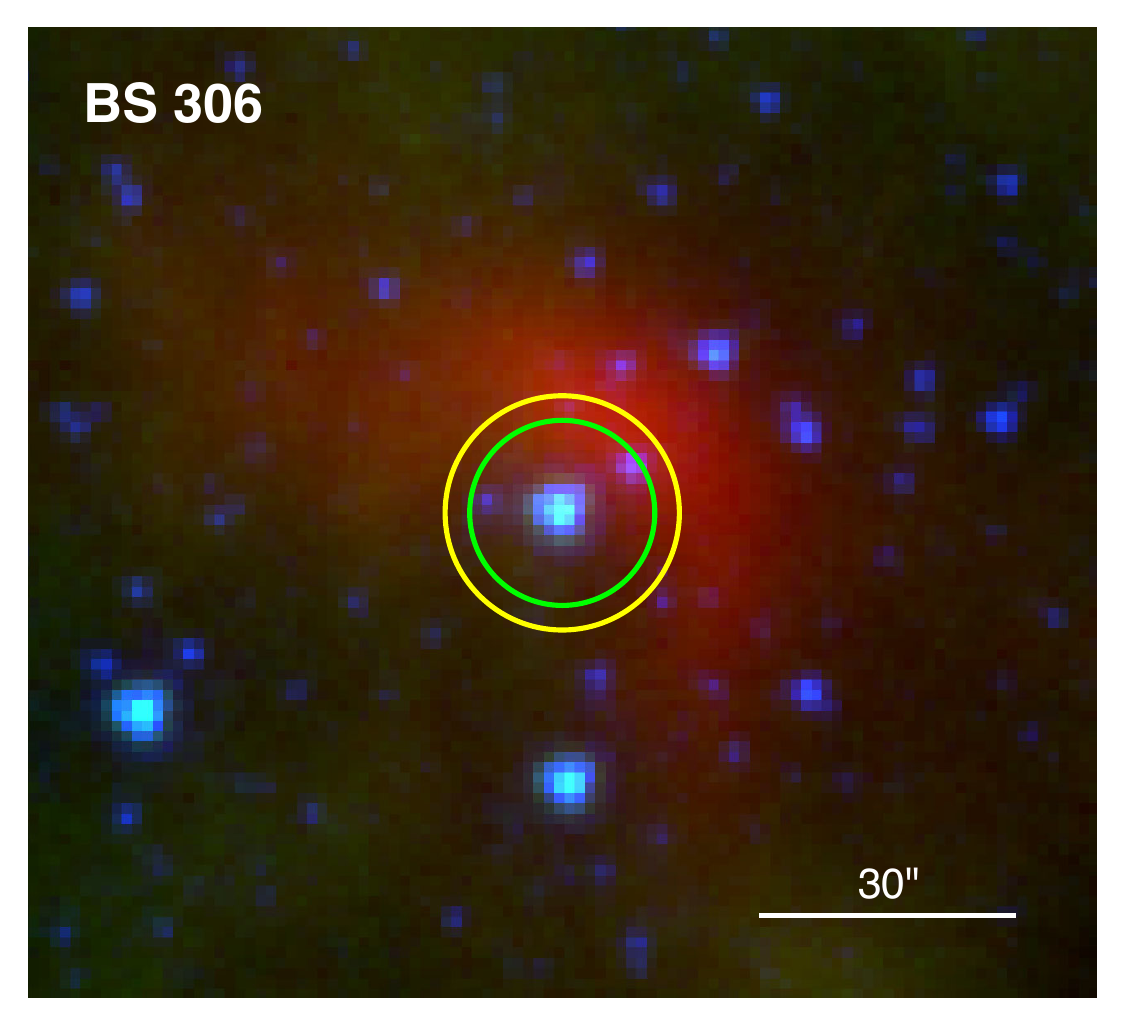}
\caption{Examples of the visual review of $R_0$ measurements overlaid on FITS mosaics from  GLIMSPE and MIPSGAL. All images are registered to Galactic coordinates, and all are zoomed to a common size indicated by the scale bars. Circles are centered on the driving star and have radii equal to various measured $R_0$ values:  white/red = classifications by individual citizen scientists indicating ``Yes!''/``Not really'' on the final classification question; yellow = $R_0$(CS) the sigma-clipped mean value of all citizen science classifications; and green = $R_0$(pre), the previous measurement by an individual ``expert'' scientist.
\label{fig:cs-master-vis_comp}}
\end{figure}

As a subjective check on the quality of the MOBStIRS citizen science measurements, we  visually inspected the $R_0$ measured for dozens of individual IR bow shocks on the original FITS survey images, using SAOImage ds9. Example visualizations of four bow shocks are shown in Figure \ref{fig:cs-master-vis_comp}. In these images we overlaid circles of radius $R_0$, to represent standoff distances irrespective of position angle. We included all individual citizen science classifications (color-coded by the response of the user to the measurement confidence question), the final citizen science value (yellow), and the previous value (green).
When inspecting IR bow shocks individually in Figure \ref{fig:cs-master-vis_comp}, we see that the citizen science and expert measurements frequently agree (bow shocks 150, 640, and 641). Where we identify a discrepancy for smaller bow shocks, for example BS 306, the MOBStIRS measurements are typically larger, as evident in Figure~\ref{fig:r0comp}, but in nearly all cases we would prefer the MOBStIRS citizen science value for $R_0$.
We conclude that the individual measurements from previous work fall generally within the distribution of citizen science measurements, albeit frequently returning values within the lower half of the range. This illustrates a key strength of citizen science, eliminating biases of individual human measurements via aggregating results from many different people.

\subsection{Asymmetric Bow Shocks}\label{sec:asymm}
Visual examination of IR bow shocks reveals that a great many of these objects deviate from the idealized, axisymmetric shapes predicted by theory \citep[][TH18]{wilkin1996,cox2012}. In particular, the IR arcs are often asymmetric, meaning the projected distance from the driving star is different toward the left or right wings. In some cases, one wing may be so faint that it disappears completely.
To identify and quantify such asymetries among the MOBStIRS sample, we compared the citizen science measurements of $R_{90}$ for the original and flipped image versions each bow shock. 
We computed the percent difference between these two $R_{90}$ values for each of the IR bow shocks as
\begin{equation}\label{eq:asym}
    \delta_{R_{90}} = \frac{R_{90, original} - R_{90, flipped}}{(R_{90, original} + R_{90, flipped}) / 2} \cdot 100.\nonumber
\end{equation}

\begin{figure}[ht!]
  \centering
  \begin{minipage}[b]{0.49\textwidth}
    \includegraphics[width=\textwidth]{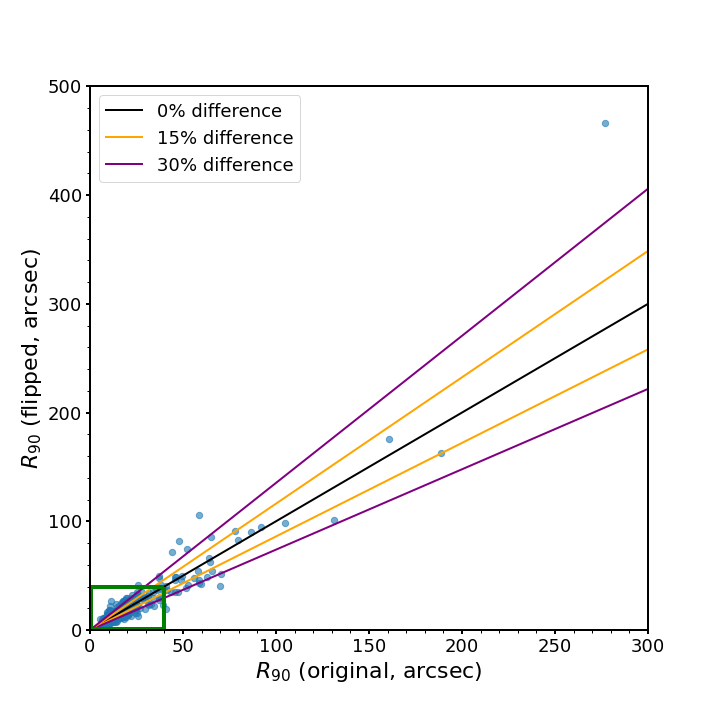}
  \end{minipage}
  \hfill
  \begin{minipage}[b]{0.49\textwidth}
    \includegraphics[width=\textwidth]{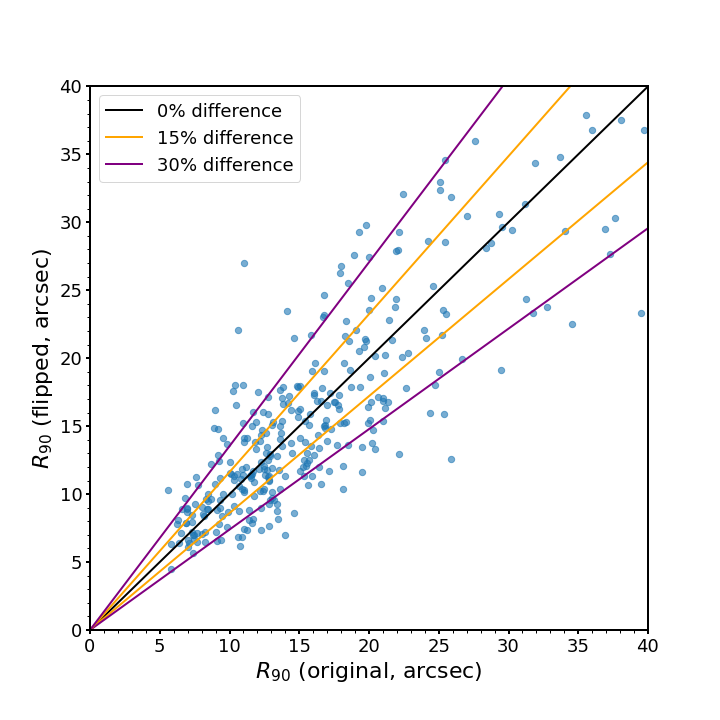}
  \end{minipage}

\caption{Plots comparing the values of $R_{90}$ for the original and flipped versions for the 361 IR bow shocks with both measurements (flag O+F = 2 in Table~\ref{tab:data}). The colored lines show the labeled values of $|\delta_{R_{90}}|$ calculated using Equation~\ref{eq:asym}. As in Figure~\ref{fig:r0comp} first plot has axes spanning the full range of data, with a green box showing the area zoomed in for the second plot.
\label{fig:r90comp}}
\end{figure}

The MOBStIRS sample includes 361 bow shocks with available measurements of $R_{90}$ in both directions (flagged as O+F = 2 in Table~\ref{tab:data}) with an average percent difference of  $\delta{R_{90}}=20\%$. This exceeds the average ${\sim}15\%$ statistical uncertainty on individual $R_{90}$ measurements, indicating that most IR bow shocks are at least marginally asymmetric.
Plots comparing the original and flipped $R_{90}$ measurements are shown in Figure \ref{fig:r90comp},
revealing a large dispersion in the deviation between the two among the MOBStIRS sample. Colored lines indicate two-sided percent differences of $|\delta_{R_{90}}| = 15\%$ and 30\%. 
A significant fraction have $\delta_{R_{90}} > 30\%$ a statistically significant deviation when compared to $R_{90}$ measurement uncertainties. We flag these as ``asymmetric" IR bow shocks, indicated by Asym. = 1 in Table \ref{tab:data} (Asym. = 0 otherwise). IR bow shocks with only a single $R_{90}$ measurement from either the original or flipped image version (O+F = 1) are also considered likely asymmetric, but are not separately flagged in the ``Asym." column, as no values of $\delta_{R_{90}}$ can be computed for them.

\begin{figure}[ht!]
  \centering
  \begin{minipage}[b]{0.49\textwidth}
    \includegraphics[width=\textwidth]{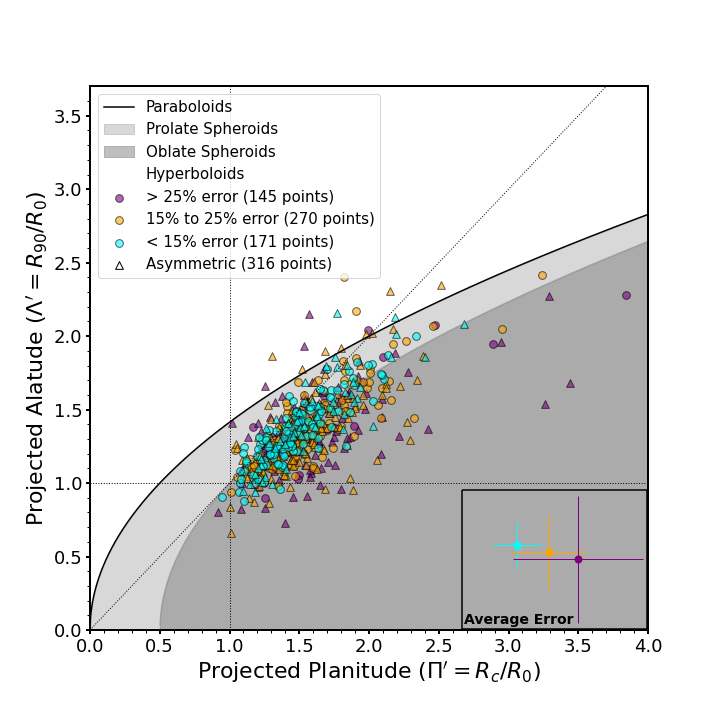}
  \end{minipage}
  \hfill
  \begin{minipage}[b]{0.49\textwidth}
    \includegraphics[width=\textwidth]{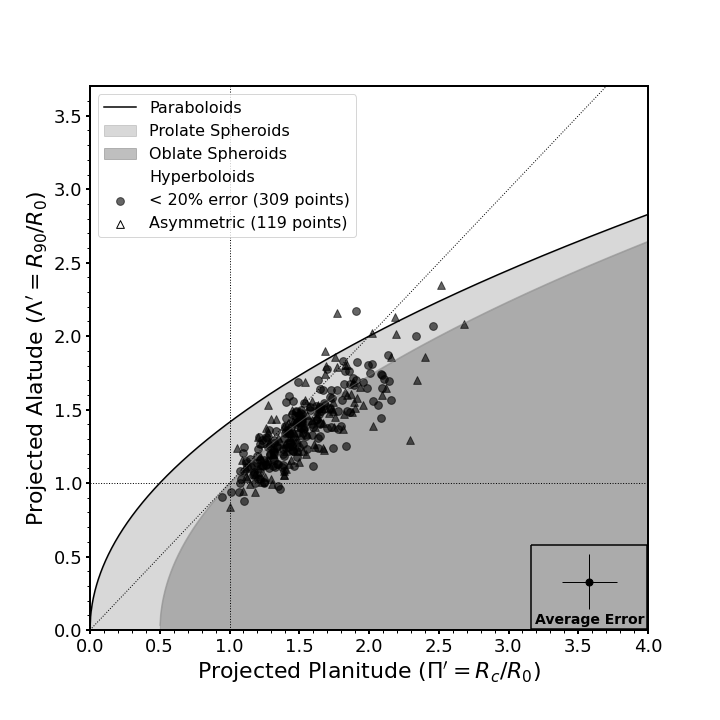}
  \end{minipage}
  \hfill
  \begin{minipage}[b]{0.49\textwidth}
    \includegraphics[width=\textwidth]{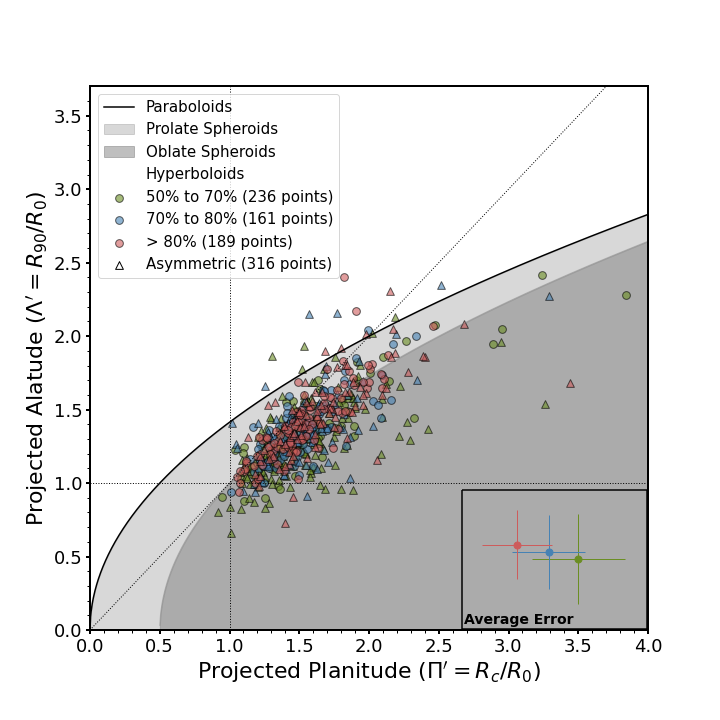}
  \end{minipage}
  \hfill
  \begin{minipage}[b]{0.49\textwidth}
    \includegraphics[width=\textwidth]{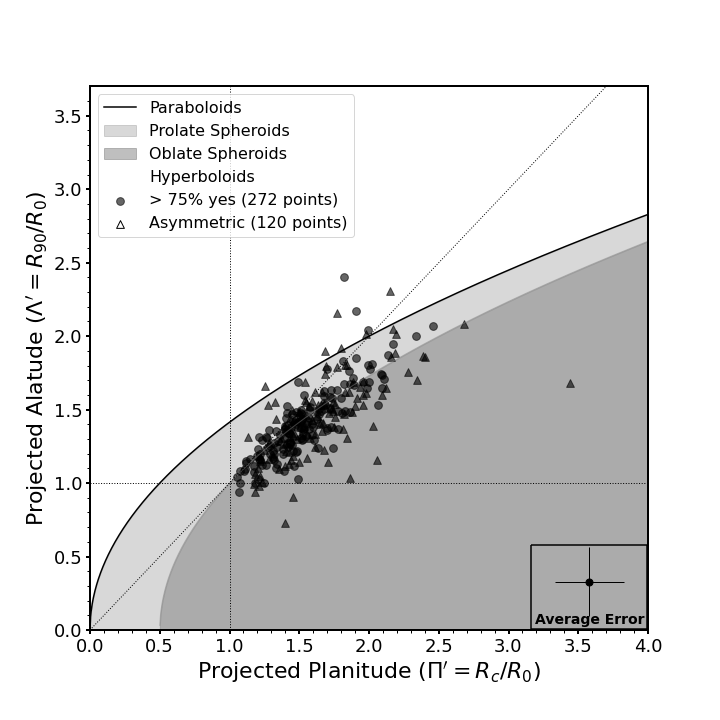}
  \end{minipage}
  \hfill

\caption{Plots of the MOBStIRS citizen science measurements on the Alatude-Planitude parameter plane defined by TH18. 
The triangles represent asymmetric bow shocks.  {\it Top left:} symbols colored according to the percent uncertainty in planitude or alatude, whichever was larger. 
{\it Top right:} Only points with fractional uncertainty less than the average (20\%) of the full dataset are plotted. 
{\it Bottom left:} symbols colored according to the percentage of citizen scientists who indicated they were confident with their measurements for a particular bow shock. 
{\it Bottom right:} Only points for which  ${>}75\%$ of citizen scientists indicated they were confident with their measurement for a particular bow shock are plotted. 
\label{fig:results}}
\end{figure}

\subsection{3-Dimensional Morphology}
Following TH18, 
we created diagnostic plots of $\Lambda'$ versus $\Pi'$, shown in Figure \ref{fig:results}. 
 The MOBStIRS sample populates a diagonal locus  with $1< \Pi' < 2.3$ and $0.9<\Lambda'< 1.8.$ This locus roughly parallels the curve denoting perfect spherical geometry, which divides oblate (dark grey) from prolate (light grey) spheroids. The locus is centered within the projected oblate spheroids region, and its spread is similar to the average measurement uncertainties.
 Asymmetric bow shocks (triangles) dominate the outliers and should be treated with caution.

\section{Discussion} \label{sec:discussion}


\subsection{Inferred Bow Shock Geometries} \label{subsec:geom}
TH18 demonstrated that the family of bow shock geometries predicted by analytical calculations of wind-wind or wind-stream momentum balance \citep[wilkinouts, cantoids, and ancantoids][]{wilkin1996,canto1996,wilkin2000,cox2012} \rev{are mostly contained within the region of projected prolate spheroids on the alatude--planitude plane} (shaded light grey in Figure~\ref{fig:results}). Hydrodynamic (HD) simulated models that include radiative cooling, and optionally incorporating magnetic fields (MHD models), create more swept-back bow shock wings \citep{acreman16,meyer17}. These more realistic simulated bowshocks have lower projected alatude values. TH18 showed that the family of HD models from \citet{meyer17} appear as marginally prolate spheroids. The alatude-planitude distribution of MOBStIRS bow shocks (Figure~\ref{fig:results}) overlaps with the locus of HD models, revealing a general agreement between the HD calculations and observed bow shock geometries. 

\rev{Compared to HD models, the \citet{meyer17} MHD models produce more compact IR arcs, implying $R_0$ smaller by as much as a factor of 2 (their Fig.\ 10, and also Fig.\ 26 of \citealt{truevsapparent}). Such a reduction in standoff distance would imply that calculated mass-loss rates could be underestimated by as much as a factor of four (Equation \ref{eqn:massloss}).}
The MHD models overlap with the HD models for all except nearly edge-on inclinations\footnote{By``edge-on'' we mean that the line segment denoting the ``true'' standoff distance, $R_0$ lies in the plane of the sky. \citet{meyer17} and TH18 define this as $i=0$, while other authors measure inclination as the angle of this line from the line-of-sight, hence edge-on is $i=90\arcdeg$ \citep{acreman16,masslossmethod,k19}.}, for which the projected planitude increases significantly, moving more deeply into the oblate spheroid region (far-right of the dark grey area in Figure~\ref{fig:results}). \rev{Only a handful of MOBStIRS bow shocks, and none with high-quality (low uncertainty, symmetric) measurements fall within the region ($2.5<\Pi'\le 6$, $\Lambda'<2$). }

Bow shocks inclined less than $30\arcdeg$ from edge-on should be the easiest to identify as mid-IR arcs, \rev{while those inclined ${>60 \arcdeg}$ might not be recognizable as arcs in our MIR images. This suggests that up to half of observed bow shocks should have inclinations placing them  within the region of Alatude-Planitude space (Figure~\ref{fig:results}) where the \citet{meyer17} MHD models diverge from the HD models.  These MHD models appear inconsistent with the distribution of morphologies among the MOBStIRS bow shocks. Recent MHD models by \citet{mackey25} explore a wider range of (weaker) magnetic field strengths and orientations from parallel to perpendicular to the bow shock symmetry axis, and these show no significant impact of magnetic fields on physical $R_0$.} 
This provides reassurance that magnetic fields
\rev{are unlikely to significantly alter standoff distances.} 

The centroid of the MOBStIRS distribution falls within the region of oblate spheroids, suggesting a marginal tension (given our relatively large measurement uncertainties, Figure~\ref{fig:results}) with the HD models plotted by TH18, which trace the high-alatude upper envelope of MOBStIRS that appear as prolate spheroids. TH18 chose to investigate \citet{meyer17} models for bow shocks produced by a runaway O star moving at $V_a=40$~km~s$^{-1}$ relative to the ambient ISM. However, most of the MOBStIRS bow shocks are expected to be produced by OB stars moving more slowly, $V_a\sim 15$~km~s$^{-1}$ \citep{k22}. More recent, 3D HD models by \citet{baalmann22} investigated the effects of varying $V_a$. Perhaps counterintuitively, the high-density wings of bow shocks produced by stars moving at high velocities are relatively wide, while lower velocities produce progressively narrower, more swept-back wings.  Narrower wings would lead to the systematically lower alatude values observed among our sample.

\subsection{Impact of Standoff Distance and Inclination on Derived Mass-Loss Rates} \citet{masslossmethod} assumed a ${\sim}10\%$ precision on $R_0$ for their mass-loss rates.
The 12.5\% measurement uncertainty averaged across the MOBStIRS sample agrees well with this previous estimate and indicates that uncertainty on $R_0$ contributes $25\%$ statistical uncertainty to mass-loss rates, on average (Equation \ref{eqn:massloss}). We recommend using the statistical uncertainty values for individual bow shocks (Table~\ref{tab:data}), as these vary greatly across the sample, ranging from 1\% to 40\%.

The distribution of the MOBStIRS sample on the projected alatude--planitude plane (Figure~\ref{fig:results}) reveals a wide range of observed inclinations. Previous work on mass-loss rates assumed an average inclination of $\langle i\rangle=65\arcdeg$ \citep[ relative to the line-of-sight;][]{masslossmethod,k19} leading to a geometric correction of $1/\sin{65\arcdeg}=1.10$ when computing physical values of standoff distance from projected $R_0$. This correction is very similar to the average statistical uncertainty on $R_0$, so it represents a significant source of systematic error. 

Theoretical work, however, presents a far more nuanced and complicated picture of possible relationships between projected and physical $R_0$ \citep[][TH18]{acreman16}. Depending upon both the underlying bow shock geometry and observational constraints of resolution and wavelength, as inclination increases relative to edge-on orientation the observed $R_0$ can be smaller, larger or  remain essentially unchanged from the true standoff distance. 

In their examination of the \citet{meyer17} models, TH18 showed that the apparent value of $R_0$ increased relative to the true value as inclination increases, the opposite of the simple geometric correction.
This occurs when the visible bow shock arc traces the limb-brightened, rear wing of the nebula, which departs from the point closest to the driving star. The choice to examine 60~\um\ simulated images with higher spatial resolution than achievable by current observatory facilities accentuated this effect.

 \citet{acreman16} presented simulated images degraded to observable spatial resolutions at 12~\um\ and 22~\um. At these shorter IR wavelengths, the bow shock morphology is typically dominated by a bright region surrounding the point closest to the driving star (the true standoff distance), where the dust is most strongly heated by stellar radiation.
As observed spatial resolution decreases, the limb-brightened arc and bright central spot merge, which tends to pull the observed $R_0$ values back toward the true standoff distance. \citet{acreman16} demonstrated that effects of limb brightening and spatial blurring precisely cancel for their 22~\um\ simulated images, which provide the best proxy for the MOBStIRS measurements at 24~\um. While the width of the 22~\um\ brightness profile measured along the symmetry axis of the arc increases as inclination departs from edge-on, the location of the peak brightness does not vary with inclination.

The above arguments suggest that inclination does not produce a significant systematic effect on measurements of $R_0$. A correction for inclination should not be necessary when computing mass-loss rates when $R_0$ has been measured using Spitzer/MIPS 24~\um\ or WISE 22~\um\ images of IR bow shocks at heliocentric distances of ${>}1$~kpc.


\subsection{Implications of Observed Asymmetries}

A majority (316/586) of MOBStIRS bow shocks are classified as asymmetric (Section~\ref{sec:asymm} and triangle points in Figure~\ref{fig:results}), of which 91 had $R_{90}$ measured toward both wings, while the remaining 224 had $R_{90}$ that could only be measured toward one wing. 

Anisotropic winds are one possible source of asymmetric bow shocks. In the case of a star that is stationary with respect to the ambient ISM and a wind outflow that is enhanced in one direction, a symmetric bow shock is produced with an ancantoid shape (TH18) that is not easily distinguishable from the case of an isotropic wind produced by a moving star. In the plausible scenario of a moving star combined with an anisotropic wind, in general there would be  misalignment between $\vec{V}_{a}$ and the direction(s) of enhanced wind flow, resulting in an asymmetic bow shock. Ancantoids (both symmetric and asymmetric) can produce lower values of apparent alatude and planitude, which would tend to drive them toward in the lower-left end of the MOBStIRS locus in Figure~\ref{fig:results}.

Density inhomogeneities (or clumping) in the ambient ISM could also produce asymmetries in observed bow shocks. Because the physical sizes of the MOBStIRS bow shocks are typically smaller than 1 pc \citep{k19}, this mechanism would require significant ISM clumping on sub-pc scales.

\rev{MHD models of bow shocks \citep[e.g.,][]{mackey25} show that both time-dependent dynamical instabilities and misalignments between the local ISM magnetic field and the star-bow shock axis can also introduce asymmetries to the shock fronts.}

\rev{The above possible causes of bow shock asymmetries are not mutually exclusive,} and our measurements are unable to distinguish between the \rev{various} scenarios. Asymmetric bow shocks should therefore be used with caution for mass-loss rate measurements, because of the potentially higher impact of unknown systematics affecting one or more factors in Equation~\ref{eqn:massloss}.



\section{Conclusions}\label{sec:conclusion}

We created the MOBStIRS website on the Zooniverse citizen science platform to measure IR bow shock shape parameters on a set of 1528 Spitzer survey images. These measurements were made primarily by several hundred undergraduate students in astronomy courses. These student citizen scientists were able to make reliable measurements for 586 individual IR bow shocks, and these measurements are of  similar or better quality to those made previously by an individual expert astronomer.

We have presented measurements for three projected shape parameters of IR bow shocks, including the standoff distance $R_0$, a key parameter for mass-loss rate measurements. We present statistical uncertainties on all of these parameters, which fall within the ranges of 12.5\% -- 15\%. Our average uncertainty for $R_0$ of 12.5\% indicates an average uncertainty of 25\% contributed by $R_0$ in the mass-loss rates of various IR bow shock-driving stars, however the uncertainty values vary widely among individual objects. A systematic correction for viewing angle does not appear necessary, as the projected $R_0$ values measured should be very close to the true standoff distances \citep{acreman16}.

The projected morphologies of the MOBStIRS sample agree well with predictions from \rev{(M)}HD simulations of IR bow shocks \citep{acreman16,mackey25}, and further appear consistent with the majority of the bow shock driving stars moving at slower than runaway speeds with respect to the ambient ISM \citep{chick20,k22}. 
The majority of IR bow shocks are classified as asymmetric, which could indicate anisotropic winds, small-scale density enhancements in the ambient ISM, \rev{and/or time and magnetic-field dependent MHD instabilities in the shock fronts}. These objects should be treated with caution in mass-loss rate determinations.

\begin{acknowledgments}
\rev{We are grateful to the anonymous referee for constructive suggestions that improved the discussion and interpretation of these results.}
We thank all of the students at Cal Poly Pomona and the University of North Texas who participated in the project, as well as the Summer 2023 University of Wyoming Research Experiences for Undergraduates cohort. We also thank Tharindu Jayasinghe for helpful conversations aiding the development of our analysis pipeline, and Dylan Hope for assisting with the visual review of citizen science measurements of individual IR bow shocks.
This work is supported by the National Science Foundation under collaborative awards AST-2108349 (Cal Poly Pomona) and AST-2108347 (U. Wyoming). This publication uses data generated via the Zooniverse platform, development of which is funded by generous support, including a Global Impact Award from Google, and by a grant from the Alfred P. Sloan Foundation. This work is based on observations made with the Spitzer Space Telescope, which was operated by the Jet Propulsion Laboratory, California Institute of Technology under a contract with NASA.

\facilities{Spitzer}
\software{SAOImage DS9 \citep{ds9}, Astropy \citep{astropy:2013, astropy:2018, astropy:2022}}
\end{acknowledgments}

\bibliography{mwp_MOBStIRS}{}
\bibliographystyle{aasjournal}



\end{document}